\documentclass[11pt,preprint]{aastex}
\usepackage{graphics}
\usepackage{amssymb}
\usepackage{amsmath}
\usepackage[dvips]{color}
\newcommand{\kms}{{km~s$^{-1}$}}
\newcommand{\fex}{[\ion{Fe}{10}]$\lambda$6376}
\newcommand{\fexi}{[\ion{Fe}{11}]$\lambda$7894}
\newcommand{\fexiv}{[\ion{Fe}{14}]$\lambda$5304}
\newcommand{\fevii}{[\ion{Fe}{7}]$\lambda$6087}
\newcommand{\arxiv}{[\ion{Ar}{14}]$\lambda$4414}
\newcommand{\sxii}{[\ion{S}{12}]$\lambda$7612}
\newcommand{\oii}{[\ion{O}{2}]$\lambda$3727}
\newcommand{\oiis}{[\ion{O}{2}]$\lambda$3729}
\newcommand{\nii}{[\ion{N}{2}]$\lambda$6583}
\newcommand{\niis}{[\ion{N}{2}]$\lambda$6548}
\newcommand{\sii}{[\ion{S}{2}]$\lambda$6716}
\newcommand{\siis}{[\ion{S}{2}]$\lambda$6731}
\newcommand{\oiii}{[\ion{O}{3}]$\lambda$5007}
\newcommand{\oiiis}{[\ion{O}{3}]$\lambda$4959}
\newcommand{\ergs}{{erg~s$^{-1}$}}
\newcommand{\target}{{SDSS J0748+4712}}
\newcommand{\jj}{{SDSS J0952+2143}}

\shorttitle{SDSS J074820.67+471214.3}
\shortauthors{Wang et al. 2010}
\received{2010 July 24}

\begin{document}

\title{Transient Super-strong Coronal Lines and Broad Bumps in the Galaxy SDSS J074820.67+471214.3}
\author{Ting-Gui Wang\altaffilmark{1,2}, Hong-Yan Zhou\altaffilmark{1,2,3},
        Li-Fan Wang\altaffilmark{4}, Hong-Lin Lu\altaffilmark{1,5},
        Dawei Xu\altaffilmark{6} }
\altaffiltext{1}{Key Laboratory for Research in Galaxies and
Cosmology, The University of Sciences and Technology of China
(USTC), Chinese Academy of Sciences, Hefei, Anhui 230026, China;
~twang@ustc.edu.cn} \altaffiltext{2}{Center for Astrophysics, USTC,
Hefei, Anhui 230026, China} \altaffiltext{3}{Polar Research
Institute of China, 451 Jinqiao Road, Pudong, Shanghai 200136,
China} \altaffiltext{4}{Department of Physics, Texas A \& M, College
Station, TX 77843, USA} \altaffiltext{5}{Physics Experiment Teaching
Center, USTC, Hefei, Anhui, 230026,China} \altaffiltext{6}{National
Astronomical Observatories, Chinese Academy of Sciences,
                 20A Datun Road, Chaoyang District, Beijing 100012, China}

\begin{abstract}
Variable super-strong coronal emission lines were observed in the
spectrum of one galaxy, SDSS J095209.56+214313.3, and their
enigmatic origin remains controversy. In this paper, we report
detection of variable broad spectral bumps, reminiscent of the 
spectrum of type II-Plateau (II-P) supernova taken a few days after 
the shock breakout, in the second galaxy with variable super strong 
coronal lines, SDSS J074820.67+471214.3. The coronal line spectrum 
shows unprecedented high ionization with strong \fex, \fexi,
\fexiv, \sxii\ and \arxiv, but without detectable optical [Fe~VII]
line emission. The coronal line luminosities are similar to thos e
observed in bright Seyfert galaxies, and 20 times more luminous than
that reported in the hottest Type IIn SN 2005ip. The coronal lines 
($\sigma \sim 120-240$ \kms) are much broader than the narrow emission 
lines ($\sigma \sim 40$ \kms) from the star forming regions in the 
galaxy, but at nearly the same systematic redshift. We also detected 
a variable non-stellar continuum in optical and UV. In the follow-up 
spectra taken 4-5 years later, the coronal lines, SN-like feature, 
and non-stellar continuum disappeared, while the \oiii\ intensity 
increased by about a factor of ten. Our analysis suggests that the 
coronal line region should be at least ten light days in size, and 
be powered either by a quasi-steady ionizing source with a soft 
X-ray luminosity at least a few $10^{42}$ \ergs or by a very luminous 
soft X-ray outburst. These findings can be more naturally explained 
by a star tidally disrupted by the central black hole than by an SN 
explosion. The similarity of their coronal line variability trend 
observed in the two galaxies suggests that the two transient events 
are of the same origin with SDSS J074820.67+471214.3 being caught at 
an earlier stage by the spectroscopic observation.

\end{abstract}

\section{Introduction}

Superstrong coronal lines have been detected in two galaxies
spectroscopically observed in the Sloan Digital Sky Survey (SDSS,
York et al. 2000), SDSS J095209.56+214313.3 (hereafter SDSS
J0952+2143, Komossa et al. 2008) and SDSS J124134.25+442639.2
(Gelbord et al. 2009). Follow-up observations of SDSS J0952+2143
showed that high ionization coronal lines were fading on time scales
of years since their discovery (Komossa et al. 2008,
2009)\footnote{We obtained two-epoch follow-up spectroscopic
observations of SDSS J124134.25+442639.2 using the MMT 6.5m
telescope at Feb. 11 and Mar. 30, 2008, and found that it shows the
similar variability trend as SDSS J0952+2143 (Zhou et al. in
prep.).}. Two physical processes were proposed for the transient
coronal lines: (1) an episodic accretion onto a supermassive black
hole (SMBH) following the tidal disruption of a star in the vicinity
of the central black hole; and (2) a peculiar type supernova (SN)
with strong coronal lines, such as SN2005ip discovered recently
(Smith et al. 2009). In the former case, it provides a way for a
systematic search for the tidal disruption events through optical
spectroscopic surveys, and thus, can be used to constrain the
stellar distribution and dynamics in the vicinity of supermassive
black holes, as well as the circum-nuclear environment. The latter
case is also interesting in that, though coronal line emission in
SNs can be fairly well understood on the basis of current models,
none of such models has been tested in these extreme cases.

Narrow coronal lines up to \fevii, \fex, \fexi, and \fexiv~were
observed in the late phase of a few SNe  (e.g., Gr\"oningsson et al.
2006; Smith et al. 2009). SN2005ip is the only SN that coronal
lines, together with recombination lines, were detected also in its
early phase, probably a few weeks after the explosion. The coronal
lines are thought to arise from the interaction between the SN and
circum-stellar material (CSM). The X-rays from the radiative shock
ionize CSM, and give rise to emission lines of a wide range
ionization states, including coronal lines. It is unclear whether
superstrong coronal emitters (SSCLEs), in which coronal lines are a
factor of several tens to hundreds more lumninous than SN2005ip, can
still be understood within this context.

Strong variable narrow coronal lines has been detected in the galactic
nucleus with soft X-ray flares (Komossa \& Bade 1999). The X-ray flares
are thought as a signature of stellar disruption by SMBH (e.g., Rees 1988;
Bade et al. 1996; Komossa et l. 2004; Halpern, Gezar \& Komossa 2004). Thus,
it is natural to associate SSCLE with the tidal
disruption process. As the tidal debris set down, it forms an accretion
disk around the black hole. The strong UV and X-ray radiation from the
accretion disk illuminates the outer debris as well as interstellar
medium further out, giving rise to broad and narrow high ionization
emission lines (e.g., Komossa \& Bade 1999).

To distinguish between the two scenarios, it is important to find
spectroscopic signatures of both of a supernova and a tidal
disruption flare. Traditionally, core-collapse supernovae have been
divided into different subtypes according to their spectral
signature and/or the shape of light-curve (see Filippenko 1997 for a
review). The different spectral signatures arise because different
stellar evolution track of SN progenitor leads to different stellar
radius, atmospheric chemical composition, and core mass. Most
massive stars will lose their entire hydrogen and even helium
envelope in Wolf-Rayet star stage, resulting in a type Ib or Ic SN;
less massive stars will retain a hydrogen shell at the end of their
evolution stage, ending with a Type II SN. Type II SN will be
further divided into Type II-L and Type II-P, depending on the lack
or presence of a plateaus in the light-curve, which may be connected
to the thickness of hydrogen shell. Stars with a mass just above
8-10 M$_\sun$ will have a thick hydrogen shell at the end of red
supergiant (RSG), and explodes as a Type II-P SN with a plateaus in
the light curve.

In this paper, we report the detection of variable broad bumps
reminiscent of a young SN II-P in the spectrum of the galaxy SDSS
J074820.67+471214.3 (\target\ for short)--a SSCLE at a redshift of
z=0.0616, which was identified during a systematic search of coronal
line emitters in the SDSS spectroscopic sample of galaxies. The
paper is organized as follows. We present the observations, data
analysis, and results in \S2. The implications on the emission line
region and ionizing continuum are discussed in \S3. We discuss
several models of SSCLEs in \S4. Throughout this paper, we assume a
$\Lambda$ dominated cosmology with $H_0=72$\kms~Mpc$^{-1}$,
$\Omega_\Lambda=0.7$ and $\Omega_M=0.3$.

\section{Observation and Data Analysis}

\subsection{Observation, Photometric and Spectral Data}

\target\ was observed by SDSS in the imaging mode at 2003-10-23. A
disk galaxy is seen in the position. An SDSS spectrum centered on
the nucleus was taken at 2004-02-20, about 4 months after the
imaging. The galaxy was detected in the GALEX survey on 2004-03-10,
20 days after the spectroscopic observation, and reobserved by GALEX
at 2010-01-09 in NUV. Four low resolution ($R\sim 600$) optical
spectra were taken with the OMR Cassegrain spectrograph mounted on
the 2.16m telescope at Xinglong station of National Astronomical
Observatory of China, one on 2008-12-25, and three on 2009-03-25
with one hour exposure each. The spectra taken at the same night are
combined. The slit is centered at the galactic nucleus with a width
of 2\farcs0 to match the seeing disk. The galaxy was detected in the
2MASS survey at 1999-02-20. The 2MASS point source catalog quoted
magnitudes of $J=16.065\pm 0.089$, $H=15.676\pm 0.132$, and
$K=15.013\pm 0.134$, with a notation of association with an extended
source. Thus, these near-infrared magnitudes can only be considered
as upper limits. All of the data were corrected for the Galactic
extinction of $E(B-V)=0.04$ (Schlegel et al. 1998) before further
analysis.

We examine the SDSS spectrum, and find three broad bumps and
numerous narrow high ionization coronal lines, including \fexiv,
\fexi, \fex, \sxii\ and \arxiv, along with other lines commonly seen
in an emission line galaxy, H$\alpha$, H$\beta$,
[\ion{N}{2}]$\lambda\lambda$6548,6583,
[\ion{S}{2}]$\lambda\lambda$6716,6732, [\ion{O}{3}]$\lambda$5007,
[\ion{O}{2}]$\lambda$3727 (Figure \ref{fig1}). The broad bumps peak
around 4050, 4600 and 6560\AA~ with widths of several hundred \AA.
These bumps do not correspond to any emission lines in an AGN, but
similar bumps are seen in the spectra of some supernovae. In the
figure, we overlay also the spectra of type II-P supernova SN1999gi
on the day of discovery (day 1; Leonard et al. 2002), which is
believed to correspond to only a few days after the shock breakout,
and another type IIP SN 2006bp, also a few days after the shock
breakout (Quimby et al. 2007). It is evident that the bumps look
similar to SN1999gi and SN2006bp although their strengths are much
larger. The broad bumps and coronal lines disappeared in the
Xinglong's spectra taken 4 and 5 years later, indicating these
features are not persistent, while [\ion{O}{3}]$\lambda$4959,5007
brightened remarkably (Figure \ref{fig2}). The similarities of the
broad bumps between \target~and SNs, including the centroids,
profiles, and transient nature, motivate us to connect both of bumps
and coronal lines to a SN-like flare.

Next, we check the spectral energy distribution (SED, Fig
\ref{fig3}). The synthesis magnitudes of the SDSS spectrum in $g$,
$r$ and $i$ bands are brighter than fiber magnitudes determined from
the SDSS imaging data by 0.51, 0.44 and 0.47 mags, respectively. The
median seeing during the spectral observation is about 2\farcs0,
similar to that used to smooth the SDSS image for fiber magnitude
estimate (Adelman-McCarthy et al. 2008), thus fiber magnitudes
properly reflect the galaxy light within the spectral fiber.
However, SDSS spectra are calibrated by tying synthesis magnitudes
to the point spread function (PSF) magnitudes for calibration stars,
as a result, synthesis magnitudes are brighter than fiber magnitudes
(Adelman-McCarthy et al. 2008). The exact difference depends on the
observational conditions and the surface brightness distribution of
the source. In this paper, we estimate the distribution of the
difference in each band using all galaxies on the same plate of
\target. These distributions are then fitted with gaussian functions
to evaluate mean deviations and their width. \target~locates far
away from the main distributions in all of the three bands. We
calculate the variations in each band by subtraction the mean
difference, and calculate the uncertainty by considering both the
width of the distribution and the uncertainty in the synthesis
magnitudes. Assuming SDSS photometric data are not affected by the
supernova, we estimate the magnitudes of SN-like event to be
19.9$\pm$0.24, 20.5$\pm$0.09 and 19.7$\pm$0.13 at $g$, $r$ and $i$
bands. At the redshift of $z=0.0616$, the absolute magnitudes are
$M_g\approx -17.3$, $M_r\approx -16.7$ and $M_i\approx -17.5$ mag.

The 2MASS photometric fluxes are lower than SDSS fluxes, indicating
that they miss part of the extended light for the galaxy. The UV
fluxes at 2004-03-10 are above the simple extrapolation of optical
SED, suggesting substantial contribution from the SN-like activity.
This is confirmed by that fact that the NUV magnitude at 2010-01-09
is about 0.68 mag fainter than that at 2004-03-10. Note that the NUV
flux at 2010-01-09 is consistent with a nature extension of optical
SED (Figure 2). Assuming that the later flux comes from the
starlight, the absolute NUV magnitude of the flare was
$M_{NUV}\approx -17.3$ mag at 2004-03-10 without considering the
intrinsic reddening. The SED of the host galaxy implies that the
contribution of the host galaxy to the FUV flux at 2004-03-10 is
much less than that in NUV, and FUV flux was dominated by the flare
component.

\subsection{Detailed Analysis of Optical Spectrum}

We fit the SDSS spectrum with the starlight from the host galaxy and a
supernova template first. The starlight is modeled as a combination of
independent components (ICs), which are convolved with a Gaussian
kernel to match the absorption line width, and shifted in redshift
to match the centroid of the absorption lines. The 6 ICs were derived
by applying Ensemble Learning ICA to the simple stellar population of
BC03 (refer to Lu et al. 2006 for detail). A uniform extinction to the
starlight is included as a free parameter.

We adopt Peter Nugent's supernova templates
\footnote{http://supernova.lbl.gov/~nugent/nugent\_templates.html}
for the SN component. The templates include 439 spectra of different
type supernovae with several ages, including Type Ia, Type Ib, Type
Ibc, Type II-L, Type II-P, and Type II-n, and peculiar type Ia
supernova, SN 1991bg and SN 1991t.  Because of similarity of
\target\ to the early spectra of SN 1999gi (II-P), SN 2006bp (II-P,
Immler et al. 2007; Quimby et al. 2007) and SN1991N (Type Ic/b,
Filippenko \& Korth 1991), we also include these spectra from
SUSPECT project \footnote{http://suspect.nhn.ou.edu } (see Leonard
et al. 2002; Quimby et al. 2007; and Matheson et al. 2001 for
original spectra). The optical spectra of the last three supernovae
were corrected for the Galactic and internal reddening from above
literature. Finally, we also added two spectra of SN 2005ip on day
1, a few weeks after SN explosion, and on day 93 after the discovery
(Smith et al. 2009). Dust extinction to the supernova component is
treated as a free parameter with the extinction curve of $R_V=3.1$
in Cardelli et al. (1989). Obvious narrow emission lines were masked
during the fit.

\begin{figure}[tbp]
\epsscale{0.8} \plotone{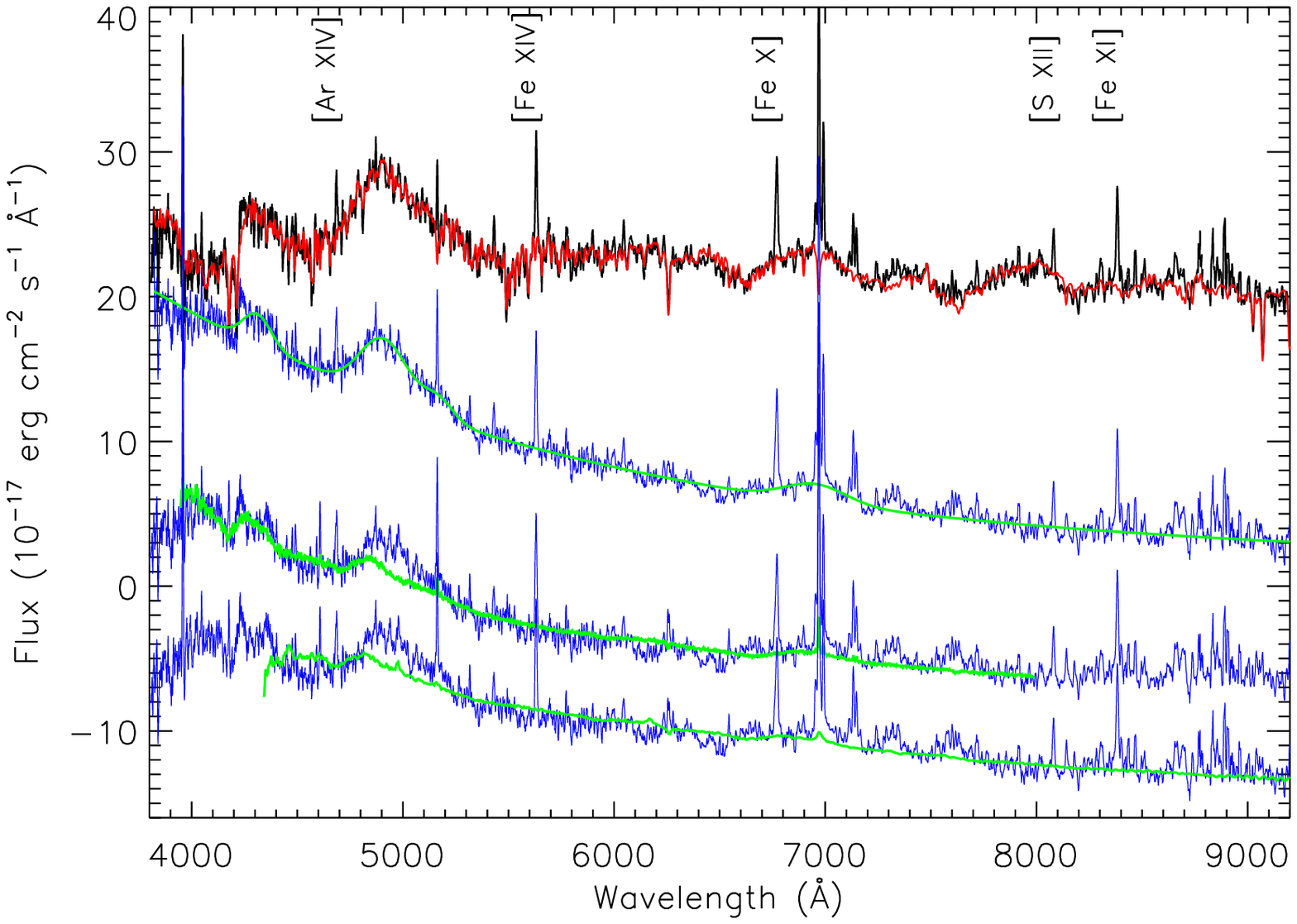} \epsscale{0.9} \plotone{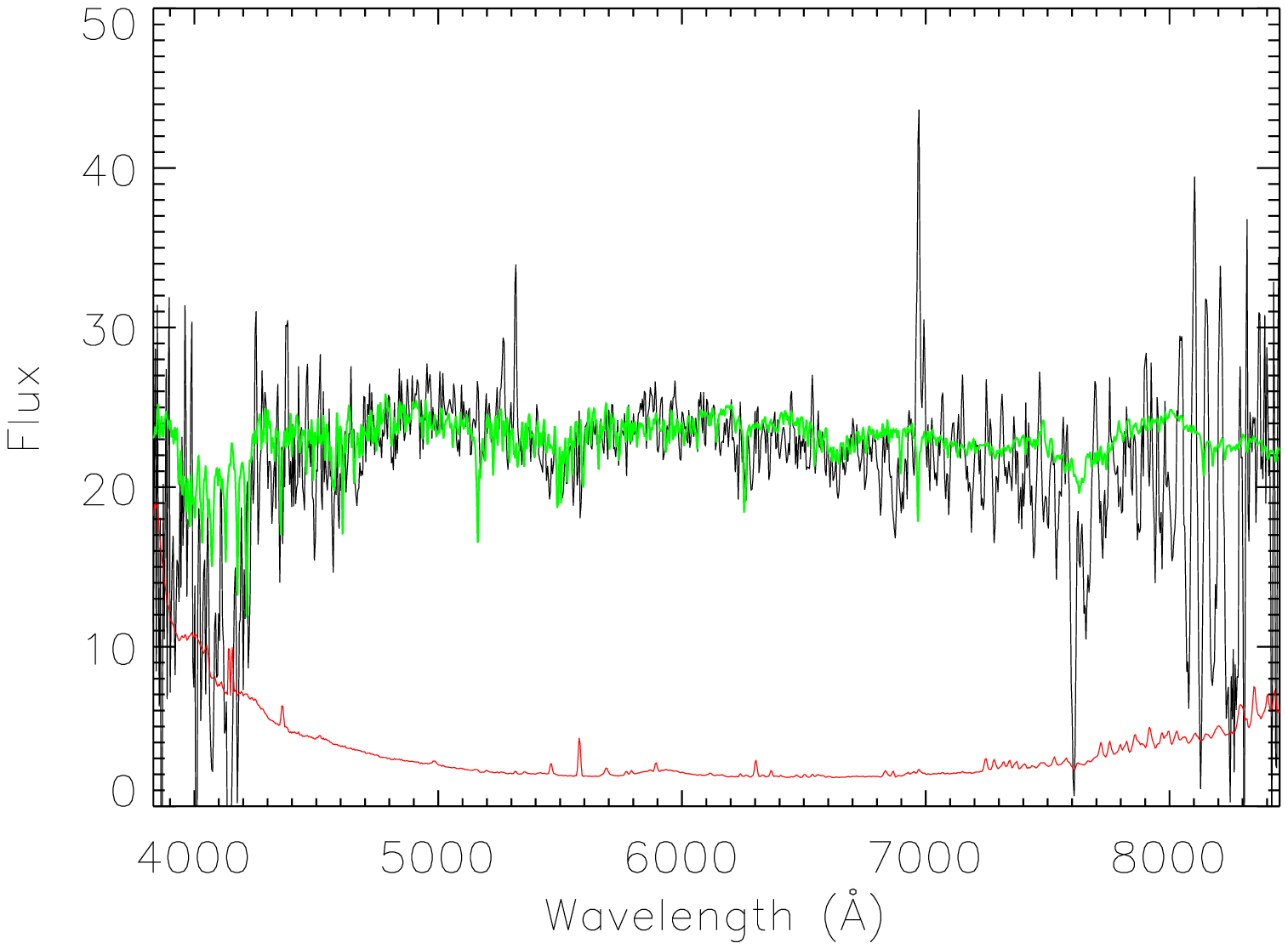}
\caption{Upper Panel:The SDSS optical spectrum (in black) of
\target\ and the best fit (in red, see text for detail). The
starlight subtracted spectrum is plotted in blue, overlaid with
different template fits (in green; models from top to bottom are
black body+Gaussian, SN1999gi on day 1, SN 2006bp on day 1, and is
shifted in vertical direction for clarity). Coronal lines are
marked. Lower Panel: The Xinglong Spectrum of \target\ and overlaid
the star-light model from fit to SDSS spectrum. The wiggle in the
red part of spectrum is due to CCD fringes. Both broad emission
lines and coronal lines disappear. \label{fig1} }
\end{figure}

\begin{figure}[tbp]
\epsscale{1.}
\plotone{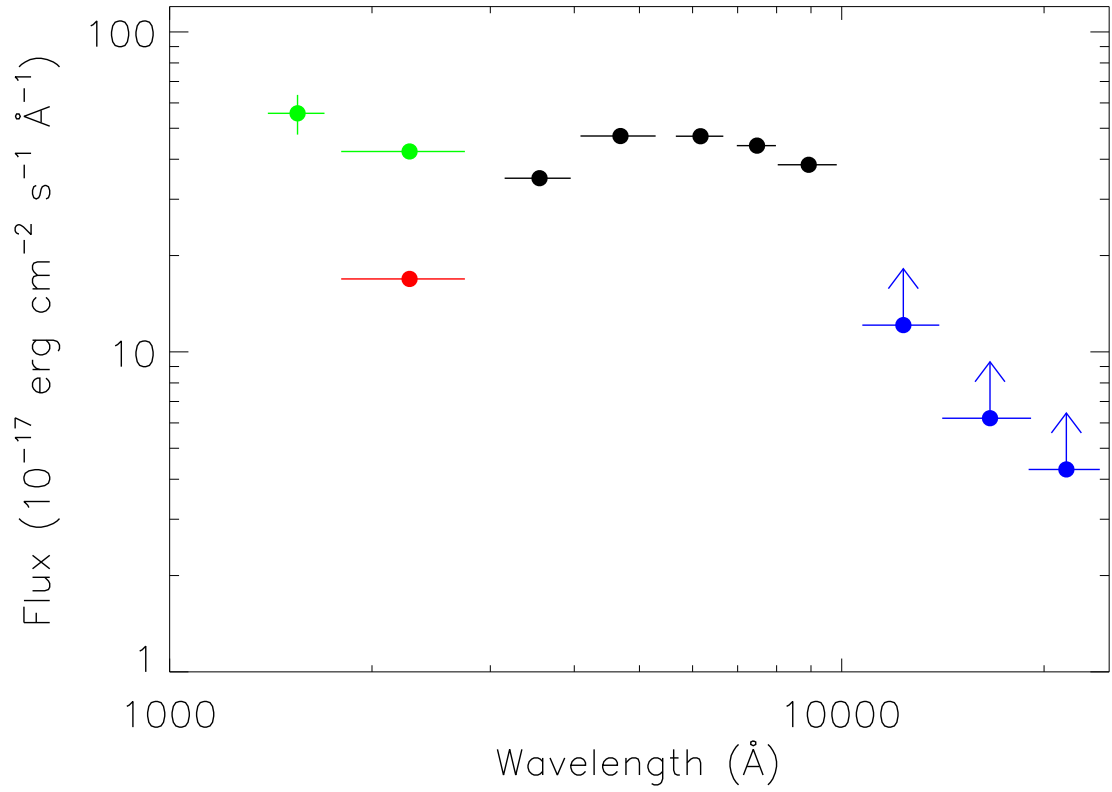}
\caption{The broad band SED of \target. The SDSS photometric data
(2003/10/23) are shown in black (model magnitudes). Near-infrared
photometric data (1999/02/20) from 2MASS point source catalog, and
UV photometric data on 2004/03/10 and 2010/01/09 from GALEX are
plotted in blue, green and red.
\label{fig2}}
\end{figure}

\begin{figure}[tbp]
\epsscale{1.} \plotone{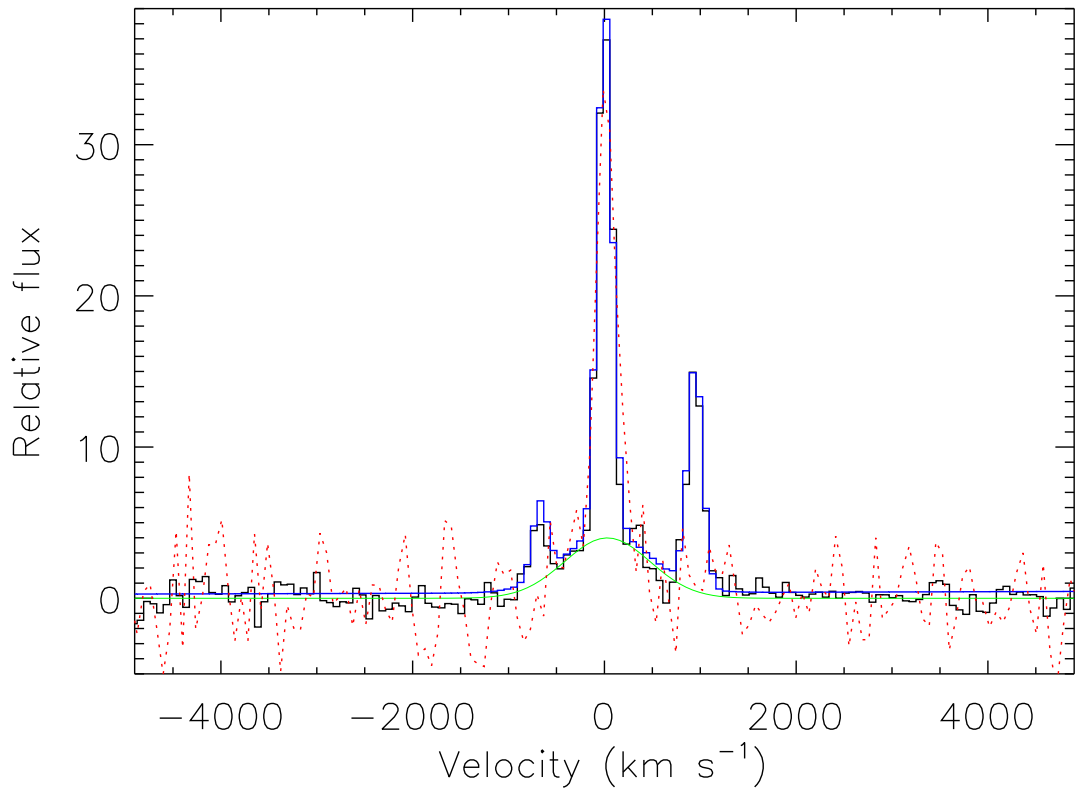} \caption{The $H\alpha$ (solid lines)
profile in the SDSS spectrum. The observed $H\alpha+[NII]$ is shown
as black, best fit as blue line, and intermediate component of
H$\alpha$ as green line. For comparison, H$\beta$ profile is
overlaid as red dashed line after scaled by a factor of 2.57. Note
that continuum and very broad component have been subtracted.}
\label{fig3}
\end{figure}

For each SN template, we obtain the best fitted parameters by
minimizing $\chi^2$. The least reduced $\chi^2$ is then sorted out
among these minima, and the corresponding supernova template is
considered as the best match. We find that SN 1999gi at day 1
provides the best fit (reduced chi-square
$\chi_\nu^2/dof=1.347/2803$), followed by SN 2006bp at day 1
($\chi_\nu^2= 1.374/2865$), also a few days after explosion. Other
templates gives significantly worse fit due to mismatch of the
bumps. As shown in Figure \ref{fig1}, these fits are still not able
to reproduce strong enough the 4600\AA~bump strength. Noticeably,
there is a global deficit in the fit from the peak to the red side
of the bump. The internal reddening of the supernova is relatively
small with $E(B-V)=0.05-0.10$ for SN199gi and SN2006bp templates.
These templates give a consistent stellar velocity dispersions
$\sigma_*\simeq 30$ \kms. Because the value is well below the
spectral resolution of SDSS and resolutions of IC templates and SDSS
spectrum are not perfectly matched, it can only be considered as an
indication for a small stellar velocity dispersion. In passing, we
note that the late spectrum of SN2005ip do not provide fit to the
SDSS spectrum although the first day spectrum of SN2005ip gives a
reasonable fit up to wavelength of 4900\AA. Because the first day
spectrum does not cover the wavelengths of two blue bumps, which are
the most prominent features in the spectrum of \target, it is not
clear if \target\ looks similar to SN2005ip at early stage.

Because SN templates do not provide good fits to the SDSS spectrum,
we also considered another empirical model consisting of a black
body and four Gaussian for bumps. With this description, we obtain a
much better fit to the SDSS spectrum ($\chi_\nu^2/dof=1.159/3216$;
also Figure \ref{fig1}). The fit yields a temperature (1.3$\pm$0.4)
$\times10^4$ K and a moderate reddening $E(B-V)=0.26\pm 0.10$,
marginally larger than the SN template fit. In the following, we
will refer this fit as our baseline model.

We measure narrow emission lines in the residual spectrum after
subtraction of the model continuum. Each emission line is fitted
with one Gaussian except for H$\alpha$, which requires an additional
broad component. In the fit to \nii\ and H$\alpha$~blending, the
width and center of the two narrow components are locked in the
velocity space. Continuum is modeled either as a local constant for
isolated or close doublets, or as a linear function in
[\ion{O}{3}]+H$\beta$, H$\alpha$+[\ion{N}{2}]+[\ion{S}{2}] regimes.
This scheme yields accepted fits for all emission lines. The derived
parameters are summarized in Table \ref{table1}. Normal narrow
lines, such as \oii, \nii, H$\alpha$~ have a width around
$\sigma\simeq 40$ \kms, after removal of the SDSS instrumental
broadening of 69 \kms. Their redshifts are consistent with being the
same. \oiii~is very weak. $H\beta$~is somewhat broader, perhaps due
to including of an intermediate component, which is prominent in
H$\alpha$. The line ratios, \nii/H$\alpha$~and \oiii/H$\beta$,
locates in the HII regime on the BPT diagram (e.g., Kewley et al.
2006). Therefore, these normal narrow lines are likely dominated by
HII regions of the galaxy. The Balmer decrement
$H\alpha/H\beta=2.57\pm0.21$ is consistent with the Case-B
recombination, indicating little dust extinction to the HII regions,
in line with low extinction to starlight inferred in the last
section. The \nii/H$\alpha$~and \sii/H$\alpha$~ratios suggest half a
solar metallicity for the HII regions (Pettini \& Pagel 2004). This
is in line with the luminosity of galaxies ($M_r=-19.75$). The
intermediate width component of H$\alpha$~has a width of around
1500\kms and a flux of 2/3 of the narrow component. Such a component
may also be present in the H$\beta$ line (Figure \ref{fig3}). As
mentioned above, we did not include this component in the fit due to
low signal to noise ratio.

\begin{figure}[tbp]
\epsscale{1.} \plotone{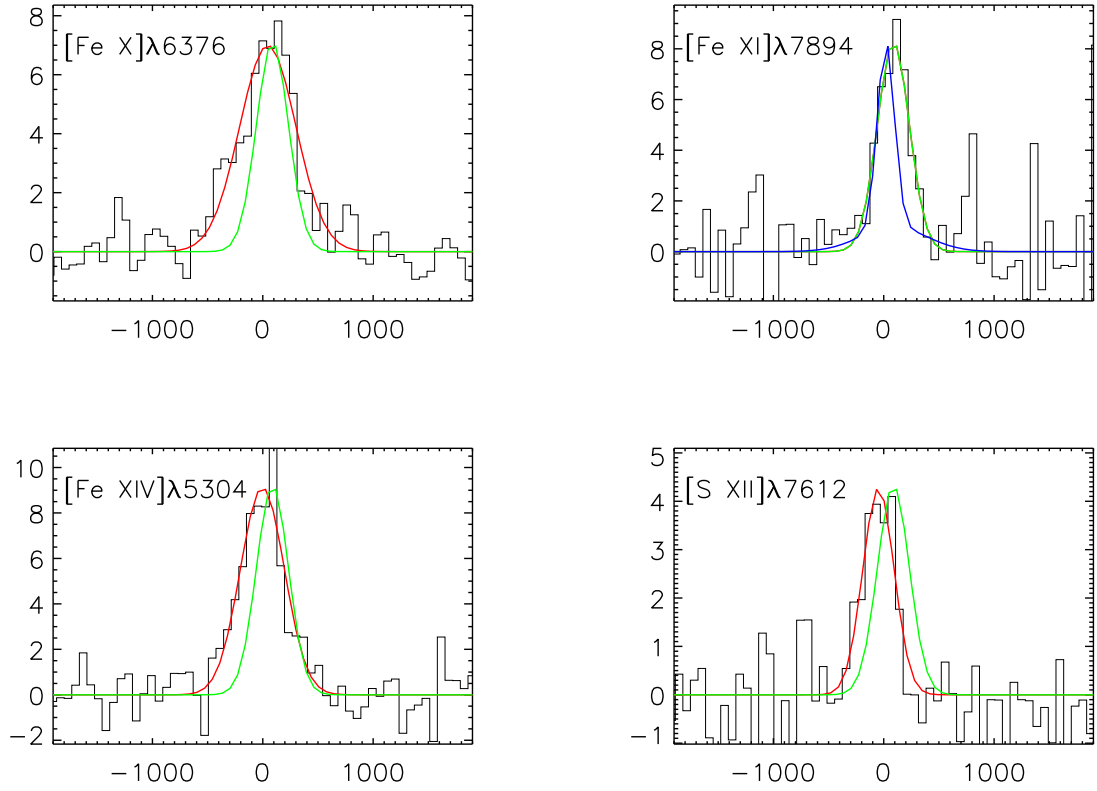} \caption{The observed profiles of
coronal lines (in black) and their Gaussian models (in red). For
comparison, the model profile of \fexi\ is over-plotted in the
Green, while in the panel for \fexi, H$\alpha$ line is shown in
blue. \label{fig4}}
\end{figure}

The coronal lines are much broader, but their profiles show subtle
difference from one line to another (Fig \ref{fig4}). \fexi\ is the
narrowest ($\sigma=127\pm16$ \kms) and redshifted by $\sim$70 \kms\
with respect to H$\alpha$\ or [NII]. It is entirely possible that
that \fex\ has a profile similar to that of \fexi. The core
component of \fex\ is very similar to that of \fexi\, but there
appears an additional weak blue component in \fex, which may be due
to contamination of other lines. If this were verified, there would
be a trend that higher ionization lines are blueshifted relative to
lower ionization lines. Other high ionization coronal lines are
blueshifted by 40 to 130 \kms\ with respect to \fexi. These
differences are larger than the uncertainty of the SDSS spectral
calibration. \fexiv\ profile is well defined, and is broader
(188$\pm$18 \kms) than and blueshifted ($-90\pm$21 \kms) with
respect to \fexi. \sxii\ is blueshifted ($-130\pm$25 \kms), but with
almost the same width (124$\pm$24\kms). \arxiv\ appears also
blueshifted and broader, though with a low S/N ratio. At a redshift
of z=0.0616, the line luminosities are (8.8, 7.4, 7.5, 3.6 and 3.5)
$\times 10^{39}$~erg~s$^{-1}$, for \fex, \fexi, \fexiv, \sxii\ and
\arxiv\ before correction for the internal reddening, which may
affect the line luminosities by no more than 50\%.

\begin{figure}[tbp]
\epsscale{1.}
\plotone{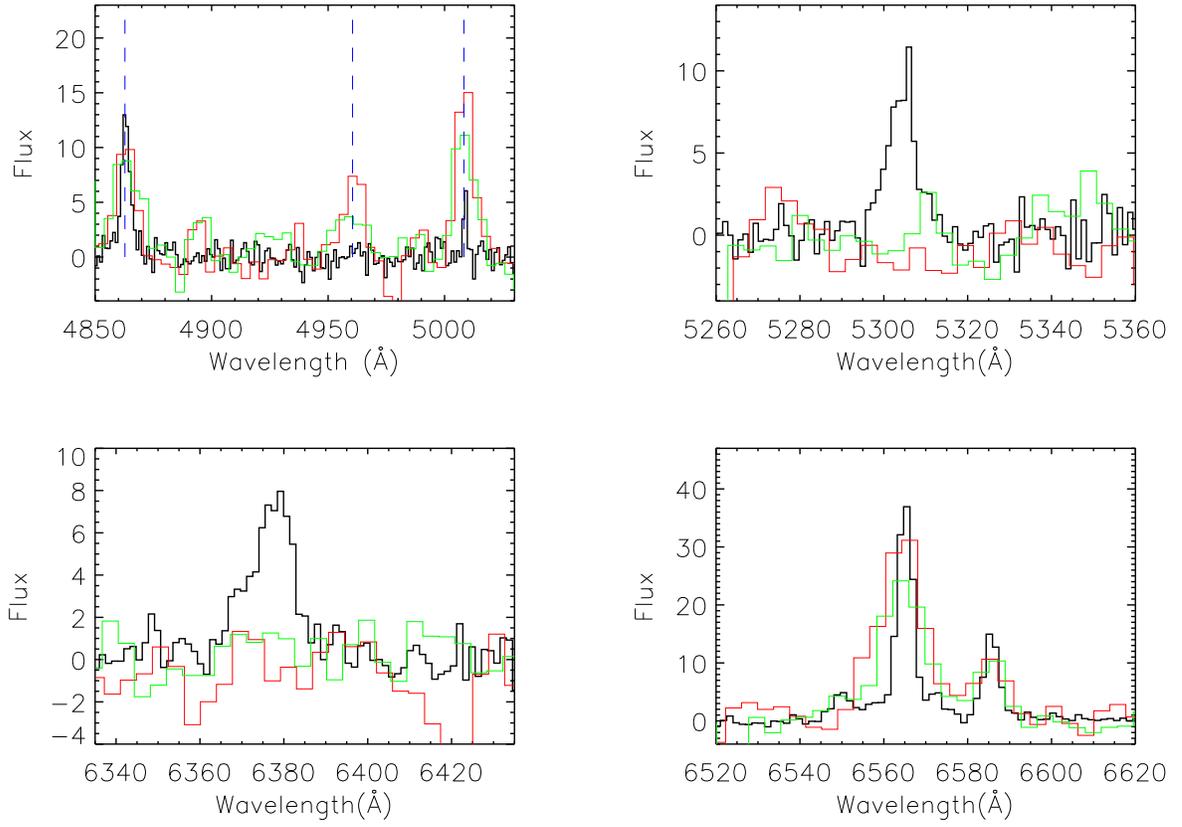}
\caption{Variations of emission lines between SDSS (black) and BAO 2.16m
observations (red , green): upper left -- H$\beta$+[\ion{O}{3}] lines; upper
right -- [\ion{Fe}{14}] line; lower left -- [\ion{Fe}{10}];
lower right -- H$\alpha$+[\ion{N}{2}]. }
\label{fig5}
\end{figure}

We also measure the narrow emission lines in the BAO spectra with a
similar procedure but without a SN continuum component. Despite the
low S/N ratio of either spectrum, the coronal lines are very weak if
present at all (Figure \ref{fig5}). However,
[\ion{O}{3}]$\lambda$5007 is 10 times stronger than in the SDSS
spectrum, suggesting the trend that low-ionization lines are being
strengthened while high ionization being weakened (Figure
\ref{fig2}).

\section{The Emission Line Region and Ionizing Continuum}
\subsection{On Coronal Line Emission Region}

\begin{figure}[tbp]
\epsscale{1.} \plotone{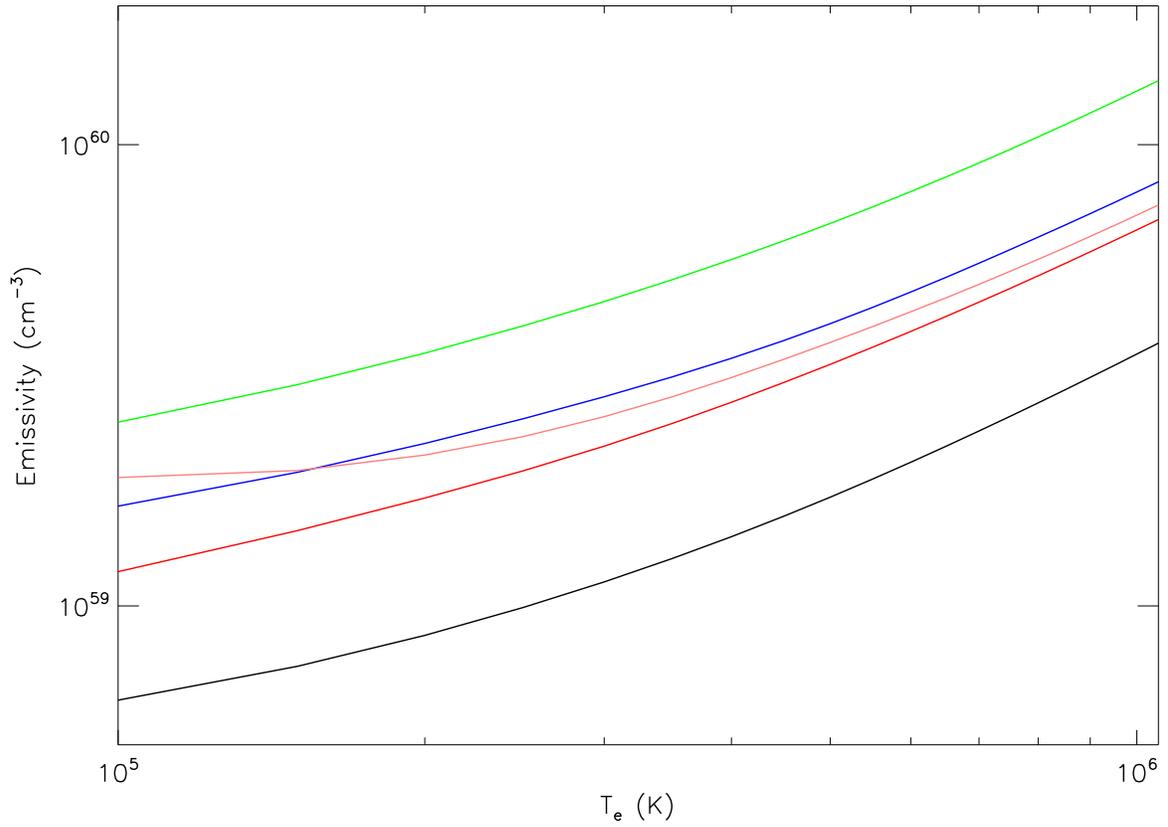} \caption{Emission measure
($n_en_{ion}V$, without internal extinction) as a function of gas
temperature for $Ar^{+13}$ (red), $Fe^{+9}$ (blue), $Fe^{+10}$
(Green), $Fe^{+13}$ (black) and $S^{+11}$ (pink).
\label{fig6}}
\end{figure}

First we estimate the Emission Measure, $EM=\int_V n_e n_{ion} dV=
<n_en_{ion}>V$, or $EM=n_e n_{ion} V$ for a uniform medium of volume
$V$, ion density $n_{ion}$ and electron density $n_e$, assuming that
collisional de-excitation and continuum fluorescence are
unimportant. The observed line luminosity relates to $EM$ with,
e.g., $L(Fe X)=C(T) n_e n(Fe^{+9}) V=C(T) EM_{Fe^{+9}}$, where
$C(T)$ is the collisional strength from the lower to the upper
levels of the correspondent transition, which is a descending
function of gas temperature $T$, and n(Fe$^{+9}$) the density of
Fe$^{+9}$. If gas is photo-ionized, its temperature is around $10^5$
K (e.g., Korista \& Ferland 1989). We take the collisional
excitation rates from the atomic data archive of CHIANTI (Dere et
al. 2009). This gives emission measures of 0.6, 2.5, 1.6, 1.9 and
1.2 $\times10^{59}$~cm$^{-3}$ for $Fe^{+13}$, $Fe^{+10}$, $Fe^{+9}$,
$S^{+11}$ and $Ar^{+13}$ using the observed line luminosities. Note
that collisional ionized gas will have a much higher temperature (a
few $10^6$~K), thus much lower $C(T)$, and requires one order of
magnitude larger $EM$ (Figure \ref{fig6}). Assuming a solar
abundance for $Ar$, i.e., $n(Ar)/n(H)=2.51\times10^{-6}$, and
$Ar^{+13}$ being the dominated species of the atom, we can derive a
minimum $EM\simeq n(H)^2V\sim 5\times 10^{64}$ [Ar/H]$^{-1}$
cm$^{-3}$. Other lines give the same order of magnitude, but
slightly lower $EM$. Considering unknown ionization correction, they
can be taken as all consistent. Since Ar, S and Fe are synthesized
only during SNe, it is unlikely that they are much more
over-abundant in circum-stellar medium unless the coronal line
region (CLR) was polluted SN process, which seems less likely (see
below).

Next we explore what constraints on CLR can be imposed based on the
above estimates. A minimum size of CLR can be set in the first
place. For a uniform spherical CLR, the radius can be written as
$R=1.1 n_9^{-2/3} EM_{65}^{1/3}[Ar/H]^{-1/3}$ lt~days, where $n_9$
is the density in units of $10^9$ cm$^{-3}$, $EM_{65}$ the emission
measure in units of $10^{65}$ cm$^{-3}$. [\ion{Fe}{10}] has the
lowest critical density of 1.5$\times10^9$ cm$^{-3}$ among these
detected coronal lines, and gas density should not be much larger
than this. This gives a minimum size of CLR of order one light day.
On the other hand, with this minimum radius, the gaseous column
density of CLR will be $N_H = 3\times 10^{24} EM^{1/3}_{65}
n_9^{1/3} $ [Ar/H]$^{-1/3}$~cm$^{-2}$. At this column density, CLR
is optically thick to X-rays up to ten keV. However,
photo-ionization requires that CLR should be optically thin to soft
X-ray rays because these photons ionize the gas. To meet this
constraint, CLR must be much larger than this minimum radius, i.e.,
having much lower density. If the column density is less than a few
$10^{23}$ cm$^{-2}$, CLR will be order of a light year and the
density around $10^6$ cm$^{-3}$.

If gas in CLR is in a thin shell or clumpy, CLR can be smaller.
In a thin shell case, the density and column density can be written as
$n_H = 7\times 10^8 EM^{1/3}_{65} \xi^{-1/2} R_{\mathrm lt\;day}^{-3/2}
$ [Ar/H]$^{-1/2}$ cm$^{-3}$, and $N_H =2\times 10^{24} EM^{1/3}_{65}
\xi^{2/3} n_9^{1/3} $ [Ar/H]$^{-1/3}$~cm$^{-2}$, where $\xi=\Delta R/R$
is the relative thickness. For a given density and $EM$, a smaller $\xi$
will result in a smaller column density and a larger radius.
If $\xi=10^{-3}$, the gaseous column density will be order of a few
10$^{22}$~cm$^{-2}$, which is still not very optically thick to soft
X-rays, and CLR is order of 10 light days.  The column density and density
for clumpy gas should be in between the thin shell and uniform sphere
case. These exercises suggest that CLR should be larger than 10 light
days in size.

Note in passing, if collisional de-excitation is important, the above
estimated $EM$ will be only a lower limit. On the other hand, Korista
\& Ferland (1989) showed that far-UV ($\sim$300\AA) continuum pumping
process accounts for nearly half of coronal emission line intensity in
Seyfert galaxies. The exact effect depends on the gas density and the
shape of ionizing continuum. For the minimum size estimated above, the
density is around $10^9$ cm$^{-3}$. At this density, collision
should dominate the excitation unless far-UV radiation field is very
strong. Thus we believe that it will not affect the order of magnitude
estimate above.

\subsection{On the Ionizing Source}

We assume that gas is photo-ionized because collisional ionization
will require much larger emissivity. The ionization potential of
Fe$^{+8}$, Fe$^{+9}$, Fe$^{+12}$, S$^{+10}$~ and Ar$^{+12}$~are
233.6, 262.1, 361, 504 and 685.9 eV, so soft X-rays are required to
ionize them. With the above estimated density, the recombination
time of the gas is $1/(n_e\alpha(T))$, typically less than 1 hour
for the ions listed above if $n_e>10^8$ cm$^{-3}$, which should be
shorter than the lasting time of X-ray radiation, while ionization
time is even shorter than this if these ions are dominated species.
Therefore, the gas is likely in quasi-ionization equilibrium. In
this case, we can estimate the ionizing photons absorbed by the
corresponding ions from their coronal lines. Each ion stays at the
upper level on average for a time of $1/\alpha_{eff}(T)n_e$ before
it recombines to a lower ionization sate, where $\alpha_{eff}$ is
recombination coefficient to excited states \footnote{Recombination
to the ground state will give rise to an ionizing photon.}. During
that period, the ion is collisionally excited $n_e
C(T)/\alpha_{eff}(T) n_e$ times. If collisional de-excitation is not
important, each collisional excitation will result in a coronal line
photon. Thus, the absorbed ionization photon flux can be written as,
taking \fex~as an example,

\begin{equation}
\Phi(FeX) \geq \frac{L(FeX)}{h\nu_{10}}\frac{\alpha_{eff,FeX}(T)}{C(T)}
\end{equation}

One can write similar equations for \fexi, \fexiv, \arxiv~and \sxii.
By summing over all terms in the left and right, we estimate that a
minimal X-ray luminosity absorbed by these ions is $\sim$ 50 of the
total coronal line luminosity. Therefore, a minimum X-ray luminosity
of a few $10^{42}$ erg~s$^{-1}$ is required to power the coronal
lines in a photo-ionization equilibrium model. This estimate is
in-exact because we have not considered two important processes.
First, recombination to lower levels will emit a photon that may be
able to ionize other ions. The re-emitted photons are distributed
nearly isotropic with an average delay of recombination time with
respective to incident ionizing radiation. Second, the ionization of
other ions requires additional X-rays. For example, when \fex~ to
\fexiv~ become dominant species, hydrogen and helium-like oxygen,
lithium-like neon, magnesium, and L-shell silicon are by up to a
factor of ten more abundant than Fe$^{+10}$, Fe$^{+11}$, Fe$^{+14}$,
S$^{+12}$~and Ar$^{+14}$ for a gas with solar metallicity. Giving
the two effects are opposite, we believe that it still gives a
reasonable order of magnitude estimate. Note that the above
estimated X-ray luminosity is much lower than the X-ray luminosity
of Seyfert galaxies with a similar coronal line luminosity ($\log
(f_{[Fe\ XI]}/f_{X})=(-3.52\pm 0.38)$ and $\log (f_{[Fe\ X]}/f_{X})
=(-3.43\pm 0.55)$; Gelbord et al. 2009).

As discussed in \S 3.1, it is quite possible that coronal line
emission sustains for years although we only have an upper limit of
4 years to the decay time. In this case, the total energy in coronal
lines would be about a few  $\sim10^{47}$~ergs. According to the
above estimate, this requires a total energy in soft X-rays of
$10^{49}$ ergs to power CLR. In addition, a large amount of UV
photons are required to ionize light elements, such as hydrogen,
helium, etc. It should be noted that coronal lines can be ionized by
an active X-ray source or they may be echoes of a past soft X-ray
flare on extended medium.

\section{The Nature of Superstrong Coronal Line Emitters}

\subsection{A Class of Superstrong Coronal Emission Galaxies?}

\target\ is the second SSCLE reported to show large line variations
on time scale of years. It shares many common properties with the
first such object, \jj\ (Komossa et al. 2008; 2009). Coronal lines
are among the strongest narrow line. \fex\ is as strong as \oiii\ in
\jj, and all high ionization coronal lines are several times
stronger than \oiii\ in \target.  An intermediate width component of
Balmer lines is present in the SDSS spectra of both objects,
although it is much stronger in \jj. In either objects, high
ionization coronal lines fades while low ionization lines remain
constant or even increases years after its discovery.  In \jj, high
ionization lines \fexiv, \fexi\ and \fex\ decreased by a factor ten
from 2005 to 2008, while [\ion{Fe}{7}]$\lambda$6087 by only 30\% and
[\ion{O}{3}]$\lambda$5007 even slightly increased; in \target,
coronal lines disappeared while [\ion{O}{3}] increased by a factor
10 from 2004 to 2009. But there are some obvious differences in
their SDSS discovery spectra. The SDSS spectrum of \target\ displays
prominent broad bumps and high ionization coronal lines but no
[\ion{Fe}{7}] and \ion{He}{2}$\lambda$4686, while the spectrum of
the latter does not show such bumps but with prominent [\ion{Fe}{7}]
and \ion{He}{2}. Also intermediate width Balmer lines are much
weaker in \target\ than in \jj. Note that \fex\ luminosity in
\target\ is a factor 5 times lower than in \jj. We find four more
SSCLEs in the SDSS spectroscopic sample of galaxies (including the
one in Gelbord et al. 2009): two without [\ion{Fe}{7}] emission
while another two similar to \jj.

The similarity between SSCLEs with and without \fevii\ indicates
that they are the same type object, while decreasing ionization with
time in both objects suggests that different ionization of the line
spectrum observed may be unified in a time evolution picture.
\target\ was observed earliest among them, so it show highest
ionization level, and also displays continuum and bump emission. As
time passes, the broad bumps disappeared and intermediate width
Balmer lines increases, and the ionization of CLR decreases. Later
on, both [\ion{Fe}{14}],[\ion{Fe}{10}] and intermediate width Balmer
lines weaken, while [\ion{O}{3}] and [\ion{Fe}{7}] strengthen.
Further on, both coronal lines and the intermediate width component
of Balmer lines disappear, and [\ion{O}{3}] and low ionization
narrow lines further brighten. The time scale for such evolution is
3-5 years. Future continuously monitoring of objects similar to
\target\ can verify this picture.

\subsection{A Type II Supernova?}

Three pieces of evidence points to a SN-like activity in \target:
0.1-0.2 magnitude brightening between the SDSS spectroscopic and
photometric observations; three broad bumps in the SDSS spectrum
similar to these seen in spectra of some young supernovae; and
vanishing of such features and coronal lines in the spectra taken 4
years later.  By comparison of SDSS photometric data with SDSS
spectrum, we estimated absolute magnitudes of the SN-like flare with
$-$16.7, $-$16.7 and $-$17.4 at $g$, $r$ and $i$ bands before
correction for internal extinctions. The internal reddening
correcting is quite modest ($<0.4$ mag in $g$). This places its
optical luminosity well in the range of a type II-P supernova
(Poznanski et al. 2009).

The broad bumps are similar to these in the spectra of Type II-P
supernovae a few days after the shock-break. In type II-P SN 2006bp,
Dessart et al (2008) identified the bump around 4000\AA~as blended
emission lines \ion{N}{3} $\lambda\lambda$4001-4099, that around
4600\AA~as blending of \ion{He}{2}$\lambda$4686,
\ion{N}{3}$\lambda$4638, \ion{C}{3}$\lambda$4647, and
\ion{H}{1}$\lambda4861$, and around 6500\AA~due to mixture of HI and
HeI. With this interpretation, the stronger 4600\AA~ bump and weak
or absent He I 5760\AA~ bump, in comparison with SN 2005ip,
indicates stronger \ion{He}{2} emission, thus a higher photosphere
temperature. Notice that the 4600\AA~bump evolves very fast in the
first few days due to rapid decrease of the photosphere temperature
(Dessart et al. 2008), and SN 1999gi, SN 2006bp are the only type
II-P known today to have been observed so early in the optical
spectra showing He II emission lines. It is unclear whether a
spectrum taken at slightly earlier time than the above two SNs will
show 4600\AA\ bump as strong as observed in this object, and whether
other type core-collapse supernovae at early phase can also
reproduce similar spectral feature. An intermediate width H$\alpha$
line is seen in the residual spectrum. Such a line is usually
considered as arising from interaction of the supernova with its
circum-stellar medium (CSM; Filippenko 1997; Pastorello et al. 2002;
Smith et al. 2010). The presence of such material can be considered
as an evidence for a massive SN progenitor, which tends to support a
core-collapse SN.

Even though a young SN II-P is a plausible explanation for the
photometric variation and the broad bumps seen in the SDSS spectrum,
such a scenario has several drawbacks in explaining coronal line
emission. First, the X-ray from young type II-P SN is far
insufficient to power the coronal lines. As noted in last section, a
total amount of energy in the soft X-rays is likely greater than
$10^{49}$ erg~s$^{-1}$. If this energy is released in the shock
breakout, which lasts for about $10^3$ s, then the X-ray luminosity
would be $10^{46}$~erg~s$^{-1}$. This is several orders of magnitude
higher than expected X-ray emission from Type II SN (e.g., Soderberg
et al. 2008; Nakar \& Sari, 2010). Alternatively, the ionizing
X-rays are produced in the shock of blast wave into the dense CSM.
However, according to Chevaliar and Fransson (1994), interaction of
expanding shell with the CSM can produce an X-ray luminosity of
order $2.4\times 10^{41} EM_{65}^{2/3}n_9^{1/3} v_{s4}^3$
erg~s$^{-1}$, assuming half of X-rays are absorbed by the shell,
where $v_{s4}$ is the shock velocity in units of $10^4$ \kms, and
$n_9$ is the particle density at shock frontier in units of
$10^9$~cm$^{-3}$ for a uniform spherical wind. This X-ray luminosity
is an order of magnitude lower than that required to power the
coronal lines if reasonable parameters are used. This energy budget
is already noted by Komossa et al. (2009).

Second, SN template fit suggests that SN is very young, a few days
after the shock breakout. If coronal line emission is related to SNe,
CLR at SDSS observing time should be less than the light travel distance
with the age of SN \footnote{In the case a continuum flare during the
SN breakout, the intense UV/X-ray ionizing photons will
ionize CSM surrounding on the way they travel out. The
ionized bubble expands nearly at the speed of light in the region
optically thin to the ionizing continuum for the ionization parameters
concerned here, but stalled after that. Due to the light traveling
effect, the observed line photons come from a thick shell around
the parabolic iso-delay surface, with a lag of SN age and a width of
continuum duration, that intersects the CSM, in the optically thin
case. In the case of $n(r)\propto r^{-2}$, taking into account of such
a delay will give a correction of order of unity.}. However, a CLR of
at least 10 light days is required to account for high coronal line
luminosity as discussed in \S 3.2.

Finally, the width of coronal lines is not consistent with a RSG
progenitor of Type II-P SN. Giving the line width of 300-450 \kms in
FWHM, it can not be formed in the shock region itself, which has a
much larger velocity, or post-shock region because the expanding
shell is opaque to these lines, rather it must be emitted from the
pre-shocked region, as in the case of SN 2005ip (Smith et al. 2009).
Also our analysis in \S 3.2 shows that CLR is much larger than the
expanding shell. Thus, the kinematics of coronal line emitting gas
reflects the undisturbed CSM. In this case, the line is broadened
due to different projected velocity of wind, and the wind velocity
must be a few hundred \kms. This requires a compact progenitor,
rather than a RSG, which produces a wind of order 15 km~s$^{-1}$. In
viewing of these problems, if supernova is responsible for the
strong coronal line emission, it must be very different from any
known SN.

\subsection{Tidal Disruption}

Tidal disruption of a star by a supermassive/intermediate mass black
hole (SMBH/IMBH) produces a flare, fading on time scales from several
months to a year, in UV and X-rays with a peak luminosity close to
Eddington one once part of debris falls back to form an accretion disk
around the black hole (e.g., Rees 1988). Systematic search in the X-ray,
UV and optical has lead the discovery of a dozen candidates (Komossa \&
Bade 2004; Halpern et al. 2004; Esquej et al. 2008; Rosswag, Ramirez-Ruiz
\& Hix 2009; Maksym, Ulmer \& Eracleous 2010; Sari et al. 2009).
Komossa et al (2008) proposed tidal disruption may be also
responsible for variable coronal line emission. Both SDSS
and our spectrum includes the galactic center, this process should be
considered as well. As Komossa et al. (2009) pointed out that tidal
disruption can potentially accounts for the observed emission line
and continuum properties of \jj\ although detail model prediction
is still lack. The UV and X-ray radiation from the accretion disk
is a natural ionizing source.  When the flare illuminates surrounding
gas, it ionizes and excites gas, giving rise to broad and narrow
emission lines, depending on the gas kinematics. The variable
intermediate-width double horn Balmer lines in that object were
interpreted as from unbounded tidal debris illuminated by the
central radiation, while coronal lines are formed by the dense
circum-nuclear gas.

Much of their arguments are valid for \target\ as well. As far as
the central back hole is greater than $10^5M_\sun$, the accretion
disk can account for high soft X-ray ionizing photons. Black hole of
$10^{5-6}M_\sun$ in the galactic center is entirely possible giving
the luminosity of the galaxy ($M_K=-22.2$), although much larger
than this can be ruled out from stellar velocity dispersion
measurement (Tremaine et al. 2002).
If CLR is virialized and dominated by the gravity of the black hole, gas
at $10^6 r_g$ will produce a line width similar around 300$-$400 km~s$^{-1}$
(FWHM), similar to the observed coronal line width. For a black hole with a
mass $10^{5-6} M_\sun$, this corresponds to a size of a few ten light days,
which meet the constraints in \S 3.2.

Photometric variations and broad bumps in the spectrum can put additional
constraints on the tidal disruption model. The accretion rate, so luminosity,
is determined by the fall-back rate of bounded tidal debris, which decreases
with time as $t^{-5/3}$ after reaching its peak (Rees 1988; c.f. Lodato, King
\& Pringle 2009). The target was brighter during the SDSS spectroscopic
observation than during the photometric observation. This set an upper limit
to the age of disruption event to 4 months during SDSS spectroscopic
observation in \target. The flare has an absolute optical magnitude
$M_g=-17.3$ during the spectroscopic observation, which is on the same
order of magnitude as predicted by some models for a $10^6M_\sun$ black
hole (e.g., Strubbe \& Quatuert 2009).

The broad bumps probably can be interpreted as optical emission lines.
Strubbe \& Quataert (2009) argued that most tidal debris should blow
away in a wind at the early times, leaving very broad optical emission
lines in the spectrum for $10^{5-6}M_\sun$ black holes. In their model,
the strongest lines are Balmer ones, which can not explain strong bumps
in the SDSS spectrum of \target. However, if a star is strongly evolved,
and much of hydrogen envelope has been stripped off. Then tidal debris
should be helium-enriched, this may explain strong blue-shifted HeII line.
It should be noted that the bumps around 4600\AA~ can be fairly well fitted
with a combination of HeII, H$\beta$ and 6560\AA~ with H$\alpha$, with their
line centers and widths locked.  Detailed physical model, which is beyond
the scope of this paper, is certainly needed to verify this explanation.

\section{Conclusion}

We detected broad bumps, reminiscent of a young type II-P supernova, and
strong high ionization coronal lines in the spectrum of \target. The
coronal line luminosity is typical of Seyfert galaxies, but
other narrow lines suggest a normal star-forming galaxy. The source
brightened by about 0.2 magnitudes in $g$-band from the SDSS imaging to SDSS
spectroscopic observations in 4 months. These bumps and coronal lines
disappeared in the spectra taken 4-5 years later, while [\ion{O}{3}] line
increases by a factor of ten. The variation trend is similar to another
SSCLE reported previously, suggesting of
the same physical origin. Their different properties between the two objects
may be ascribed to different observed evolution stage.

We set a robust lower limit on the size of CLR to be 10 light days,
and on the total energy of the ionizing continuum in soft X-rays to
be $10^{49}$~erg. Both the size of emission line region, high soft
X-ray luminosity and broad width of coronal lines cannot be
understood in the young Type II-P supernova context. If coronal
lines are indeed associated with supernova explosion, then the
supernova must be very different from that current known. We argued
that tidal disruption of evolved star by a massive black hole may
provide a viable explanation from the bumps and coronal lines. In
this model, the bumps are considered as broad emission lines from
winds produced during the tidal disrupted process (Strubbe \&
Quataert 2009), and the star is partially evolved.

A critical test to the above scenario should come from continuous
monitoring the spectral evolution and early follow-up in other
bands, such as X-ray, UV and infrared band, shortly after the
discovery of the coronal line emitter. X-ray and UV emission from
supernova predicted in current model drops very fast after hours and
ten days (Nakar \& Sari, 2010), while in the tidal disruption model,
the tidal debris is accreted on the time scale of years (e.g.,
Lodato, King \& Pringle 2009). We found that the absolute magnitude
of the flare in GALEX NUV band is -17.3 at 20 days after the SDSS
spectroscopic observation for \target, which is brighter than that
predicted by supernova models, but consistent with the tidal
disruption model. We plan to carry out such a survey with future
Chinese spectroscopic survey telescope--LAMOST, and conduct early
follow-up observations of such events with other observatories.

\acknowledgements We are grateful to the referees for throughout
reading and critical comments that lead to significant improvement
of the paper, and to Dr Nathan Smith for providing the spectra of SN
2005ip. This work was supported by the Chinese NSF through
NSF-10973013, 10973012 and 11033007, the national 973 program
2007CB815403 and 05, and CAS knowledge innovation project No.
1730812341. DX acknowledges support from the Chinese NSF under 
grant NSFC 10873017, and from program 973 (2009CB824800). This work 
has made use of the data obtained by SDSS and
by 2.16m optical telescope on Xinglong station, Chinese National
observatories. Funding for the SDSS and SDSS-II has been provided by
the Alfred P. Sloan Foundation, the Participating Institutions, the
National Science Foundation, the U.S. Department of Energy, the
National Aeronautics and Space Administration, the Japanese
Monbukagakusho, the Max Planck Society, and the Higher Education
Funding Council for England. The SDSS Web Site is
http://www.sdss.org/.


\begin{deluxetable}{llll}
 \tablecaption{Narrow Emission Line Parameters \tablenotemark{a}
 \label{table1} }
 \tablewidth{0pt}
 \tablehead{
 \colhead{Line} & \colhead{$\lambda$} & \colhead{$\sigma$\tablenotemark{b}} &
 \colhead{$Flux$} \\
\colhead{} & \colhead{\AA} & \colhead{km~s$^{-1}$} & \colhead{}
 }

\startdata

\oii      & 3727.30$\pm$0.14 & 85 $\pm$ 8   &  39$\pm$ 5 \\
\oiis     & 3730.06          & 85       &  52$\pm$5 \\
\arxiv    & 4414.28$\pm$0.55 & 246$\pm$22 &   39 $\pm$ 5 \\
H$\beta$  & 4863.16$\pm$0.15 & 122$\pm$ 9  &   61 $\pm$ 4 \\
\oiiis    & 4960.91$\pm$0.19 & 69\tablenotemark{c} & 4$\pm$ 2 \\
\oiii     & 5008.85          & 69       &  13 $\pm$ 3 \\
\fexiv    & 5304.34$\pm$0.29 & 199$\pm$17 &   80 $\pm$ 6 \\
\fex      & 6377.21$\pm$0.34 & 251$\pm$16 &   94 $\pm$ 6 \\
\niis     & 6550.29$\pm$0.05 &  80 $\pm$ 2  &   21 $\pm$ 1 \\
H$\alpha$ & 6565.10        & 80    &   152$\pm$ 6 \\
H$\alpha$ & 6566.40$\pm$0.90 & 460$\pm$59 &    99 $\pm$11 \\
\nii      & 6585.70        & 80   &   64 $\pm$ 3 \\
\sii      & 6718.80$\pm$0.14 & 79 $\pm$ 6 &    35 $\pm$3 \\
\siis     & 6733.18          & 79         &    24 $\pm$3 \\
\sxii     & 7611.87$\pm$0.90 & 142$\pm$59 &   38 $\pm$11 \\
\fexi     & 7896.34$\pm$0.37 & 145$\pm$14 &   81 $\pm$ 7
\enddata
\tablenotetext{a}{Flux in units of $10^{-17}$ erg~s$^{-1}$~cm$^{-2}$
in the observed frame.} \tablenotetext{b}{Including the SDSS
instrument width of about 69 \kms.} \tablenotetext{c}{The value is
pegged at the lower-limit during the fit.}
\end{deluxetable}

\end{document}